\documentclass{article}
\usepackage{spconf,amsmath,graphicx,xcolor}

\usepackage{multirow,subcaption,booktabs,soul,url,cite}

\title{TSOG: A format for Temporally and Spatially Ordered Gaussians}
\name{Shady Gmira, Evangelos Alexiou, Emmanouil Potetsianakis, Emmanuel Thomas}
\address{Xiaomi Technology Netherlands B.V, The Hague, The Netherlands}
\begin{document}

\maketitle
%
\begin{abstract}
We propose Temporally and Spatially Ordered Gaussians (TSOG), a format for efficient representation of 4D Gaussian Splatting (4DGS) content. 
TSOG extends the Spatially Ordered Gaussians (SOG) framework to the temporal domain by introducing a timeline attribute and temporal parameterization of geometry and appearance attributes.
Similar to SOG, TSOG is a lossy format that assigns each Gaussian a unique index and encodes attribute values as index-aligned image data. 
TSOG is model-agnostic, extensible, and compatible with both discrete and continuous 4DGS representations.
Evaluation using a PLYs sequence and FreeTimeGS as baselines, serving as simplistic and state-of-the-art 4DGS representations respectively, shows file size reductions exceeding 90\%, with PSNR differences ranging between \(-0.42\) and \(+0.85\) dB.
These results demonstrate substantial file size savings with minimal quality degradation, enabling efficient representation, storage, and delivery of dynamic scenes for next-generation 4D content. 
\end{abstract}
%

\section{Introduction}
\label{sec:intro}
Gaussian Splatting (GS) has emerged as a key topic in academic and industrial research on real-time neural rendering.
GS represents 3D scenes as a collection of unstructured, anisotropic 3D Gaussians, with each Gaussian determined by its position, scale, orientation, opacity, and color.
The ability of GS to represent 3D scenes with high fidelity and photorealistic quality has made it a compelling alternative to traditional mesh- or point-based representations. 
Kerbl et al.~\cite{kerbl_3d_2023} demonstrated that radiance fields can be rendered interactively by projecting millions of Gaussians directly onto the screen, enabling free-viewpoint, high-quality views of the 3D scene at real-time frame rates. 

More recently, GS was extended to represent dynamic scenes by adding the time dimension, leading to 4D Gaussian Splatting (4DGS).
By capturing temporal changes in geometry and appearance, 4DGS is suitable for applications in immersive media, XR, and virtual production, among others~\cite{li_spacetime_2023, wang_freetimegs:_2025, Wu_2024_CVPR, xu2024representing}.
4DGS representations can be categorized into discrete and continuous; the former represent temporal evolution at distinct time instances, and the latter model temporal evolution as a continuous function, allowing rendering at any arbitrary time.
As dynamic scene representations become increasingly relevant, efficient storage, transmission, and rendering of 4DGS content have emerged as important challenges.

For storage and transmission, most existing 4DGS content representations rely on established 3DGS file formats, such as PLY~\cite{turk1994ply}, typically involving custom adaptations to accommodate temporal data. 
Extending the PLY format with method-specific attributes hinders interoperability, whereas using sequences of PLY files (one per frame) results in excessive storage overhead and poor scalability. 
Ongoing efforts to standardize 3DGS and 4DGS file formats aim to address this gap, highlighting their importance for real-world deployment.

In this paper, we present TSOG (Temporally and Spatially Ordered Gaussians), an extension of SOG (Spatially Ordered Gaussians) 
to 4DGS content.
SOG~\cite{morgenstern_compact_2025} is a file format for 3DGS content, widely supported by rendering pipelines~\cite{lichtfeld2025, playcanvas_2026, sparkjsdev_2025},
that offers efficient access, storage, and delivery.
However, it lacks support for a temporal dimension. 
To address this limitation, we propose two key features: (a) a timeline attribute specifying each Gaussian's lifespan, and (b) temporal parameterization of geometry and appearance attributes across the Gaussian's lifespan.
The resulting TSOG format inherits the advantages of SOG, while being modular, extensible, and backward compatible. 
Crucially, it does not modify the underlying 4DGS generation method; instead, it defines a method to represent and store 4DGS content. 

Our contributions are: (i) we define the TSOG file format as a flexible extension of SOG with native temporal support; (ii) we experimentally evaluate TSOG in terms of file size and rendering quality; and (iii) we demonstrate the TSOG representation format by integrating it into PlayCanvas, highlighting its potential for wide deployment.

\section{Related work}
\label{sec:related}
Various file formats have been used to store and exchange GS content, each reflecting different design priorities such as simplicity, compression efficiency, or rendering convenience. 

A popular representation is the extended PLY format, adapted by Kerbl et al.~\cite{kerbl_3d_2023}.
Originally designed to store data from 3D scanners~\cite{turk1994ply}, PLY was extended with per-Gaussian attributes such as scale, orientation, opacity, and spherical harmonics. 
In this format, attribute values are repeated for every Gaussian, leading to redundancies and large files. 
However, due to its foundational role in the original 3DGS codebase, virtually all subsequent implementations have adopted it as the de facto output format. 
For 4DGS content, temporal information is often incorporated by further extending the attribute list with time-related parameters~\cite{wang_freetimegs:_2025, li_spacetime_2023}. 
In such cases, custom format extensions and renderers are required, limiting interoperability and adoption across platforms.

Another approach for 4DGS content representation is a sequence of PLY files (one per frame).
This approach, while straightforward, has significant limitations: large storage requirements, inefficient transmission, and poor scalability to scenes with millions of Gaussians or long temporal sequences. 
Specifically, it fails to exploit the inherent structure of many 4DGS generation methods, which typically update per-Gaussian attributes over time rather than redefining a new set of Gaussians at each frame, leading to data duplication. 
For example, a 2-second dynamic scene can require more than 5.5\,GB of data and nearly 25 million Gaussians.

The SPZ format, developed by Niantic (Scaniverse), is an open-source binary format designed for 3DGS content~\cite{nianticlabs_2025}. 
It uses aggressive quantization and a column-based binary layout optimized for GS data. 
GS attributes are stored in a fixed order using reduced-precision representations: positions use 24-bit fixed-point integers, scales and colors use 8-bit log-encoded integers or unsigned integers, orientation quaternions are stored as 3 components in 8-bit signed integers (with the fourth component derived), and spherical harmonics coefficients are stored as 8-bit signed integers. 
This compact representation, combined with gzip compression, achieves around 90\% reduction in file size compared to PLY. 
Although highly effective for static scenes, SPZ does not natively support 4DGS content and lacks extensibility.

SOG operates differently, reformulating GS attributes as image data~\cite{morgenstern_compact_2025, sog_2026}. 
Gaussians are first reordered by position using a Morton (Z-order) curve~\cite{morton1966computer} to improve spatial coherence in the resulting 2D layout, and then mapped onto a 2D grid where each pixel's position corresponds to a unique Gaussian index.
GS attributes are encoded as index-aligned images, enabling efficient compression using standard image codecs.
Redundant GS attributes, such as scales or spherical harmonics, are further compressed by using vector quantization with codebooks stored in metadata. 
This design achieves substantial file size reductions (15-20$\times$ smaller than PLY), while maintaining ease of deployment and extensibility. 
However, it lacks support for the temporal dimension. 

Finally, glTF has been explored as a container for GS data through the standardized \texttt{KHR\_gaussian\_splatting} extension~\cite{khronosgroup_2025}. 
This approach encodes GS attributes (position, scale, orientation, opacity, and spherical harmonics) as point primitive attributes within glTF's buffer structure.
However, this GS extension is tailored to 3DGS content.


\section{TSOG File Format}
\label{sec:tsog_format}
\subsection{Definition}
TSOG extends the SOG file format~\cite{sog_2026} to support 4DGS content.
In TSOG, a Gaussian is defined by the attributes: position, scale, orientation, opacity, color, and timeline; the first three denote geometry attributes and the next two appearance attributes.
The temporal evolution of a scene is represented by Gaussians whose geometry and appearance evolve along their lifespan, as determined by the timeline attribute. 
The TSOG file format retains the core SOG principle of a lossy design by quantizing and storing per-Gaussian parameters as index-aligned image data.
TSOG extends it by introducing a temporal dimension through the timeline attribute and temporal parameterization of geometry and appearance attributes, enabling its application to both discrete and continuous 4DGS representations. 
A fundamental requirement is that for every Gaussian, identified by a unique linear index $i$, its timeline and temporally parameterized 
attribute values must map to the same pixel location across all corresponding attribute images.
TSOG is defined as a representation format and does not prescribe a particular generation or rendering model. 

\subsection{Timeline Attribute}
\label{sec:timeline_attribute}
The timeline attribute comprises temporal parameters that specify the lifespan of every Gaussian within or relative to the scene's timeline, which serves as a common time axis.
This lifespan is shared across the Gaussian's geometry and appearance attributes and defines the domain of their temporal parameterization.
The temporal parameters are chosen from one of the following three options:
(1) \textit{Frame id}: Specifies the frame index in which this Gaussian is active. 
The \textit{fps} may be additionally signalled to specify corresponding time instances.
(2) \textit{Start} and \textit{duration}: \textit{Start} specifies the onset, and \textit{duration} specifies the length of a Gaussian’s lifespan.
When \textit{fps} is signalled, they represent frame units; otherwise, time instances.
(3) \textit{Center} and \textit{scale}: \textit{Center} indicates the peak visibility time, and \textit{scale} represents the extent of a Gaussian's lifespan around the peak. 
The \textit{scene length} may be included to specify the duration of the scene (e.g., in sec).
The first two options are more suitable for discrete 4DGS representations, while the last is better for continuous 4DGS representations.

In all cases, temporal parameter values are stored per-Gaussian in a newly introduced TSOG attribute image, namely, timeline attribute image, at pixel locations that are index-aligned with the other attribute images. 
In the timeline attribute image, the \textit{frame id}, \textit{start}, or \textit{center} values are mapped to the red (R) channel, and the \textit{duration} or \textit{scale} value is mapped to the green (G) channel, with corresponding entries in \texttt{mins} and \texttt{maxs} specifying their range. 
All values are preserved at 16-bit precision and stored in two separate image files (upper- and lower-significance bits) following the SOG convention.  
An example of the \texttt{timeline} structure syntax and corresponding semantics is given as follows:
\vspace{-0.2cm}
\begin{itemize} 
\item{\texttt{type}: Enumerated type of temporal parameters. If~0, \textit{frame id}.\,If 1, \textit{start} and \textit{duration}.\,If 2, \textit{center} and \textit{scale}.\nolinebreak}
\vspace{-0.2cm}
\item{\texttt{fps}: Number of frames per seconds. Optionally signalled when \texttt{type} is equal to 0 or 1.}
\vspace{-0.2cm}
\item{\texttt{mins}/\texttt{maxs}: Min/max values used to map temporal parameter values to image intensities for all channels.}
\vspace{-0.2cm}
\item{\texttt{files}: Image file paths. Images contain 8 upper (u) or 8 lower (l) significance bits. Filename convention: \texttt{timeline\_<l|u>.webp}.}
\vspace{-0.2cm}
\end{itemize}
\begin{small}
\begin{verbatim}
"timeline": {
 "type": [/* scalar value */>] 
 "fps": [/* scalar value */>] 
 "mins": [<t1_min>, <t2_min>],
 "maxs": [<t1_max>, <t2_max>],
 "files": [
  "timeline_l.webp",
  "timeline_u.webp"
 ]
}
\end{verbatim}
\end{small}

\subsection{Temporal Parameterization of Attributes}
Every geometry and appearance attribute of a Gaussian is parameterized over time and can be categorized as static (time-invariant) or dynamic (time-varying) over its lifespan, as determined by the timeline attribute.
Notably, this temporal parameterization type (i.e., static or dynamic) is consistent for a given attribute across all Gaussians in the scene.

Static attributes are stored in SOG images (i.e., defined by SOG).
Dynamic attributes are split between SOG images and newly introduced TSOG images.
For dynamic attributes modelled using polynomial temporal parameterization, base values (i.e., constant 0th-order coefficients) corresponding to the \textit{frame id}, \textit{start} or \textit{center} of a Gaussian's timeline are stored in SOG images.
Their higher-order coefficients describing their temporal evolution are stored in TSOG images specified under the \texttt{temporal} structure.
SOG and TSOG images store attribute values for all Gaussians comprising a scene. 

The \texttt{temporal} structure supports polynomial temporal parameterizations, which are individually defined per dynamic attribute. 
For each dynamic attribute, a dedicated TSOG image is used for each higher-order coefficient, with \texttt{mins} and \texttt{maxs} arrays specifying corresponding ranges.
All values are preserved at 16-bit precision and stored in two separate image files using upper- and lower-significance bits. 
TSOG images use the same channel mapping (e.g., R, G, B, A) as the SOG images.
An example of the \texttt{temporal} structure syntax and its corresponding fields is provided below:\nolinebreak
\vspace{-0.2cm}
\begin{itemize} 
\item \texttt{mins}/\texttt{maxs}: Min/max values used to map higher-order coefficients to image intensities for all channels and per polynomial degree, sorted in increasing order. 
\vspace{-0.2cm}
\item \texttt{files}: Image file paths. Images contain 8 upper (u) or 8 lower (l) significance bits. Filename convention: \texttt{temporal\_<attr>\_<order>\_<l|u>.webp}. 
\vspace{-0.2cm}
\end{itemize}
\begin{small}
\begin{verbatim}
"temporal": {
 "means": {
  "mins": [
   [<mx_min_1>, <my_min_1>, <mz_min_1>]
  ],
  "maxs": [
   [<mx_max_1>, <my_max_1>, <mz_max_1>]
  ],
  "files": [
    "temporal_means_1_l.webp",
    "temporal_means_1_u.webp",
  ]
 },
 "quats": {
  "mins": [
   [<qx_min_1>, <qy_min_1>, <qz_min_1>]
   [<qx_min_2>, <qy_min_2>, <qz_min_2>]
  ],
  "maxs": [
   [<qx_max_1>, <qy_max_1>, <qz_max_1>],
   [<qx_max_2>, <qy_max_2>, <qz_max_2>]
  ],
  "files": [
   "temporal_quats_1_l.webp",
   "temporal_quats_1_u.webp",
   "temporal_quats_2_l.webp",
   "temporal_quats_2_u.webp",
  ]
 },
}
\end{verbatim}
\end{small}
This example assumes 1st-order polynomial temporal parameterization for geometry and 2nd-order for orientation. 
The syntax can be accordingly adapted to other attributes or temporal evolution models. 

\subsection{Mapping Attributes}
The TSOG file format preserves the original SOG~\cite{sog_2026} mapping for static attributes and base values of dynamic attributes without modification (i.e., SOG images). 
Such attribute values are stored exactly as in SOG~\cite{sog_2026}, using the same spatial reordering, index alignment, channel layout, normalization, quantization, and metadata structure. 
As a result, existing SOG decoding pipelines can process these images unaltered. 

The TSOG file format extends this mapping strategy to newly introduced images storing the timeline attribute and higher-order coefficients of dynamic attributes (i.e., TSOG images). 
The TSOG images follow the same 2D grid layout and per-Gaussian index alignment as in the SOG images. 
Therefore, the same Gaussian corresponds to the same pixel location across all attribute images, enabling independent decoding while preserving attribute synchronization.
Moreover, the same normalization and 16-bit split-byte storage convention is used, ensuring the same encoding and decoding mechanisms are applied as in SOG images. 

\begin{figure}[t]
\centering
\begin{subfigure}{0.5\textwidth}
\begin{subfigure}{0.49\textwidth}
  \centering
  \includegraphics[width=\linewidth]{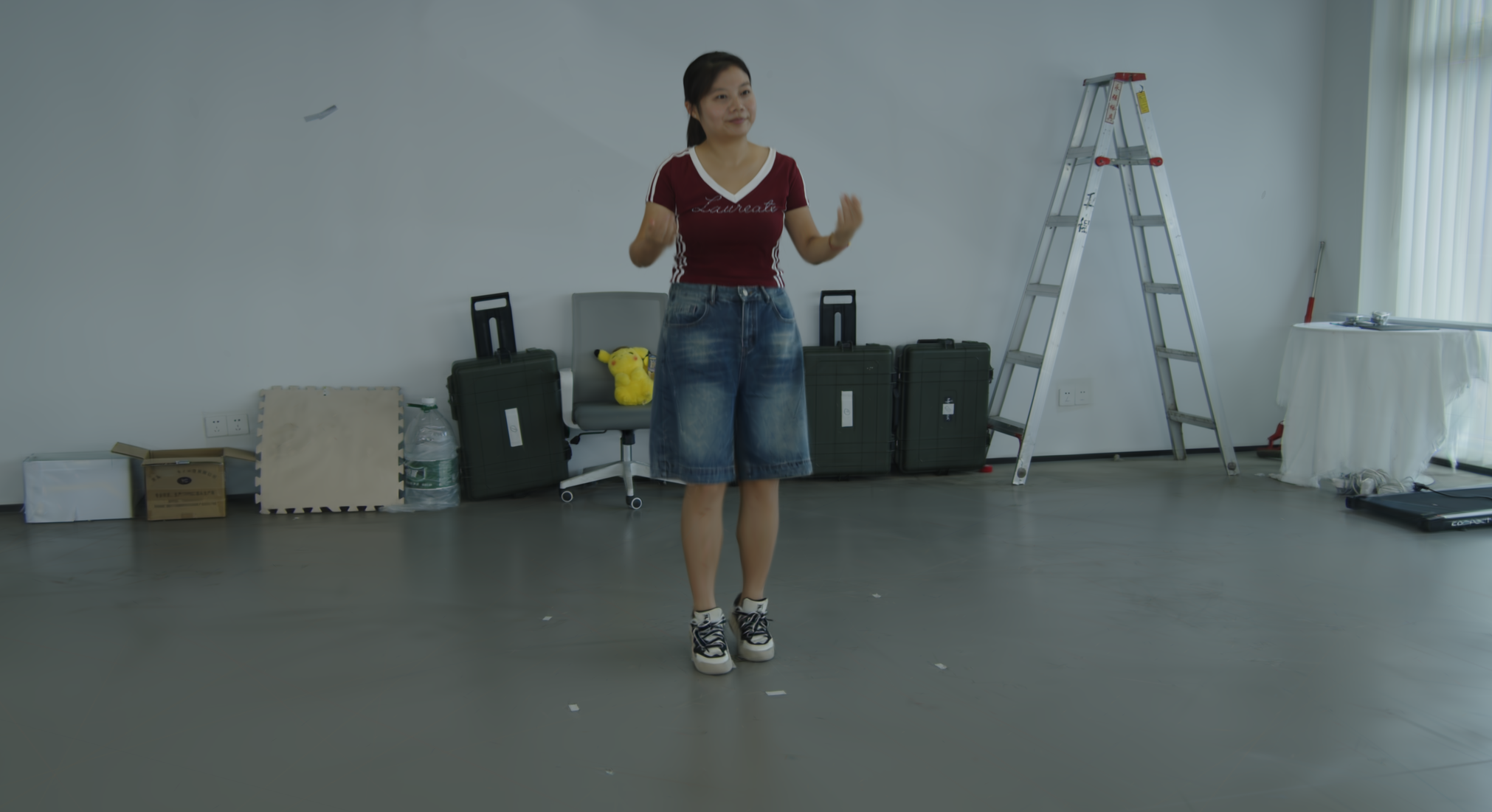}
  \caption{PLYs sequence}
\end{subfigure}
\hfill
\begin{subfigure}{0.49\textwidth}
  \centering
  \includegraphics[width=\linewidth]{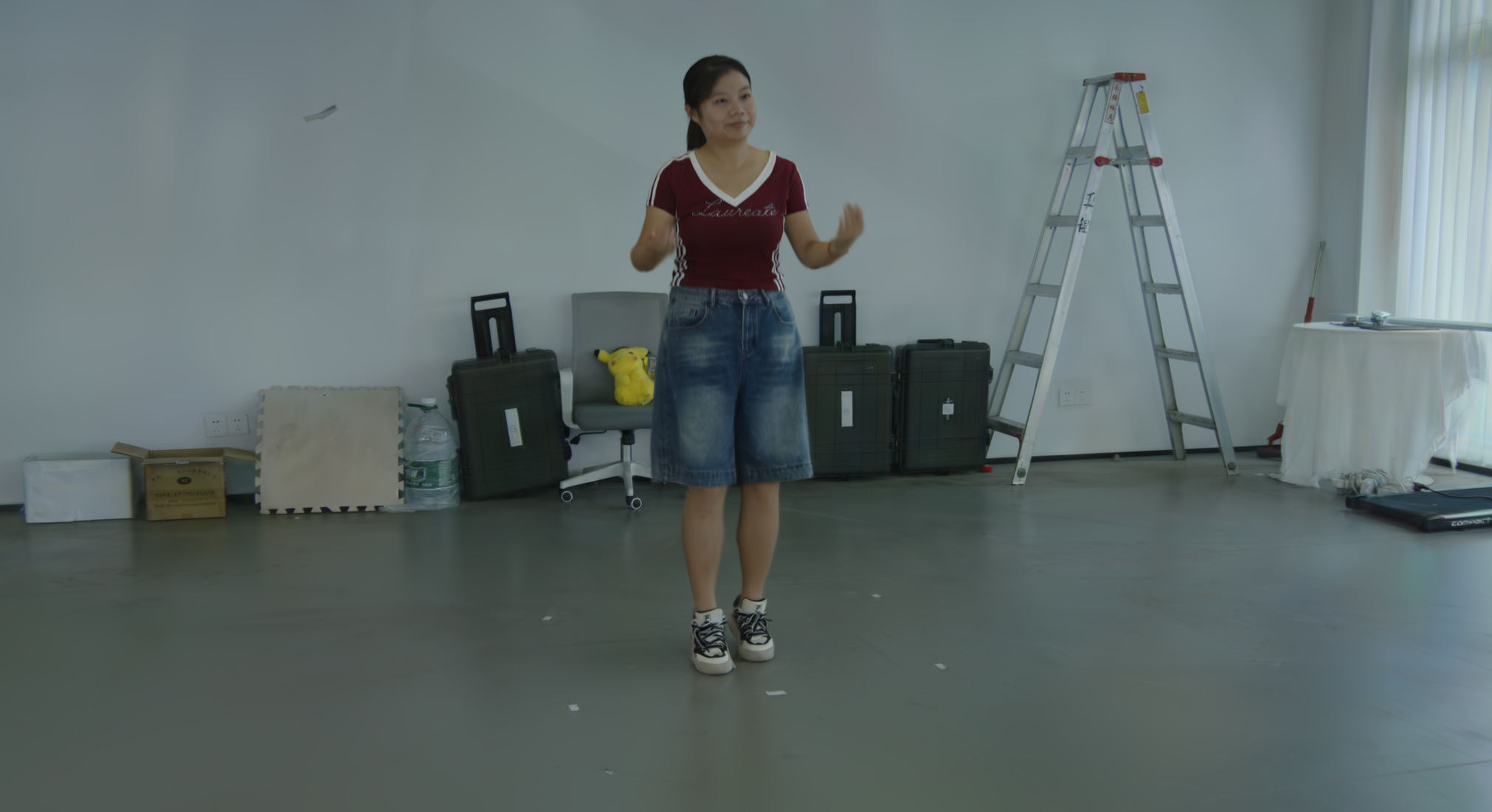}
  \caption{TSOG}
\end{subfigure}
\end{subfigure}
\caption{Qualitative comparison of PLYs sequence and TSOG, for VV sequence 001\_1\_seq0, frame \#73.} 
\label{fig:temporal_scale_comparison}
\vspace{-0.5cm}
\end{figure}

\section{Experiments}
\label{sec:exp}
\subsection{Experimental setup}
We evaluate the proposed TSOG file format in two complementary experiments using two datasets:
the dataset released for the compression track of the SIGGRAPH Asia 2025 Volumetric Video challenge~\cite{vvc2025siggraph} (VV), and the Neural 3D~\cite{neural_3d} dataset excluding Flame Salmon.
The source 4DGS contents are generated by training corresponding dynamic scene representations.
For the VV dataset, the 2 provided scenes were already trained using FreeTimeGS~\cite{wang_freetimegs:_2025}. 
For the Neural 3D dataset, due to the lack of a public implementation of FreeTimeGS~\cite{wang_freetimegs:_2025}, 
each of the 5 selected scenes is trained using FreeTimeGS\-Vanilla~\cite{ftgs_vanilla} at half resolution with a target of 5~million Gaussians at SH3. 
The output of this training step is a file per scene using an extended PLY format with FreeTimeGS-specific temporal parameters~\cite{wang_freetimegs:_2025}.

Then, for each experiment, a distinct baseline is created and used.
In the first experiment, a sequence of PLY files serves as the baseline for the most common discrete 4DGS representation. 
In this case, we evaluate the trained model at every discrete time instance (frame) and save the resulting 3D Gaussians as an independent PLY file per frame.
In the second experiment, FreeTimeGS~\cite{wang_freetimegs:_2025} serves as the baseline for an advanced continuous 4DGS representation with temporal parameterization. 
In this case, we use the original extended PLY files directly from the training step, which embed relevant temporal attributes. 
In each experiment, PSNR and SSIM are computed between images rendered from the baseline and the corresponding TSOG representation. 
All reported metrics are computed on half-resolution rendered images.
All evaluations are performed using a single desktop RTX 4090.

\subsection{PLYs Sequence Representation}
\label{ssec:plys}
In this experiment, a PLYs sequence represents the baseline, with each frame stored as an independent PLY file corresponding to a specific discrete time instance. 
To convert the PLYs sequence to the TSOG file format, all Gaussian attributes, including positions, orientations, scales, opacity, color, and timeline, are stored as static attributes. 
In addition, each Gaussian is assigned to one frame $f \in \{0,\dots,N-1\}$, where $N$ is the frame count of the sequence. 
The timeline attribute is configured with \texttt{type} set to 0, mapping the \textit{frame id} to the frame index assigned to the Gaussian, indicating that it is active only at that specific frame.


For the VV dataset, we use the first 100 frames of each scene, whereas for the Neural 3D dataset, we use the first 20 frames of each scene. 
This frame limit is imposed by the underlying WebP library used during TSOG encoding, which cannot handle images larger than roughly 25\,million pixels.
The limitation is more restrictive for Neural 3D scenes due to the larger number of active Gaussians per frame obtained using FreeTimeGS-Vanilla~\cite{ftgs_vanilla} for training (1--1.5\,million), compared to the active Gaussians per frame present in VV scenes trained using the official FreeTimeGS~\cite{wang_freetimegs:_2025} (0.2--0.5\,million).



\begin{figure}[t]
\centering
\begin{subfigure}{0.48\textwidth}
\begin{subfigure}{0.49\textwidth}
  \centering
  \includegraphics[width=\linewidth]{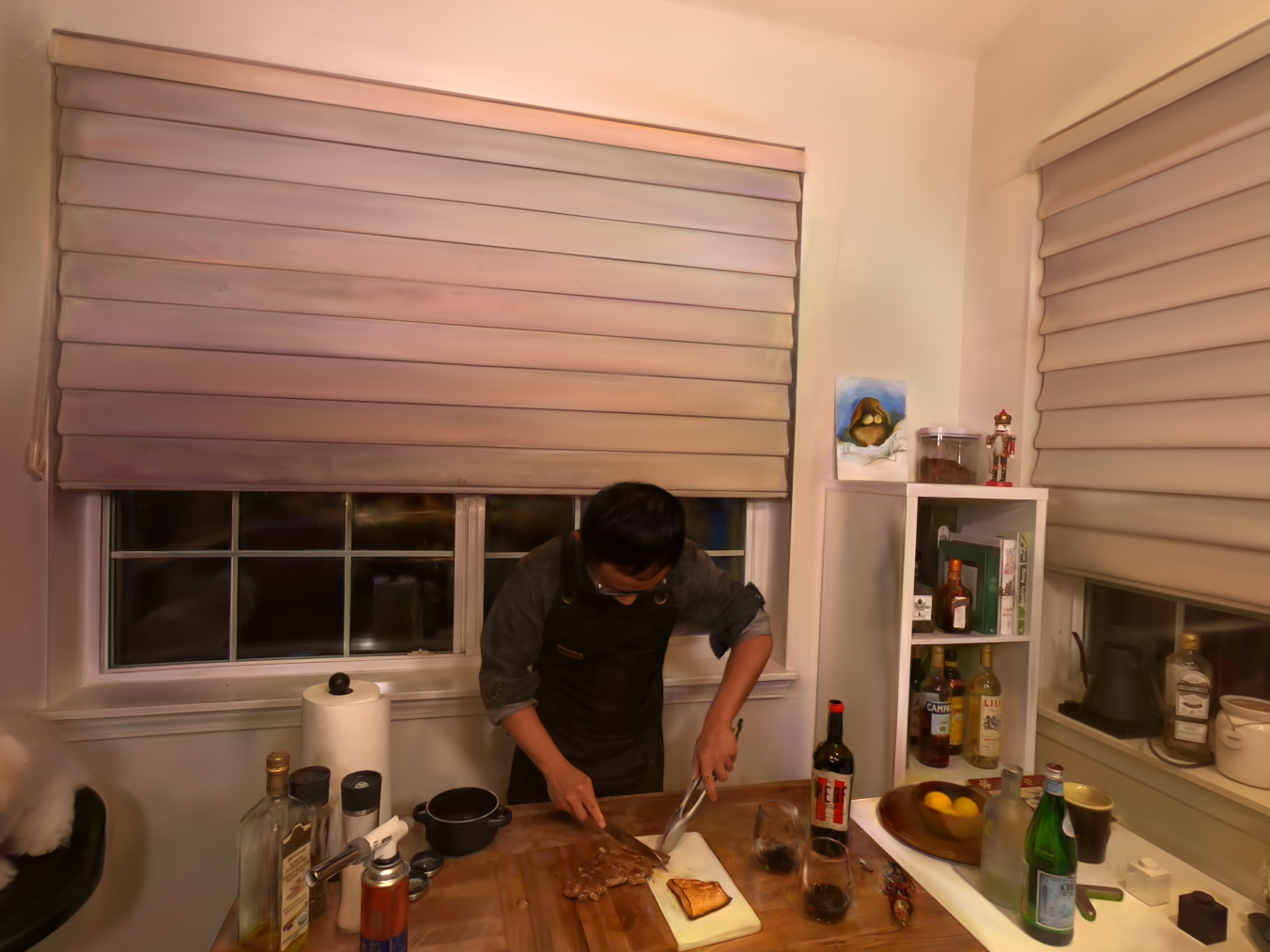}
  \caption{FreeTimeGS}
  \label{fig:q_ply_001}
\end{subfigure}
\hfill
\begin{subfigure}{0.49\textwidth}
  \centering
  \includegraphics[width=\linewidth]{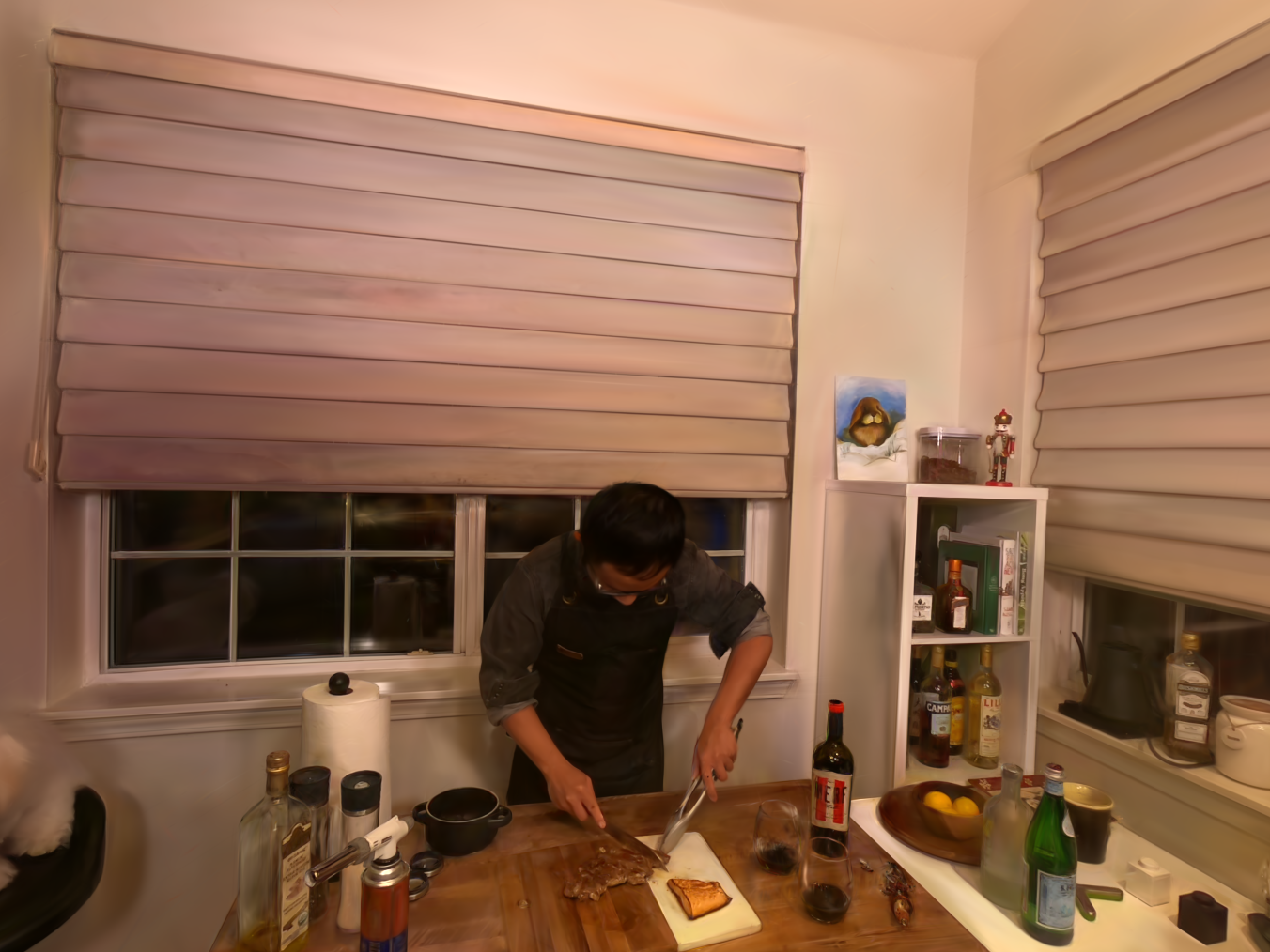}
  \caption{TSOG}
  \label{fig:q_tsog_001}
\end{subfigure}
\end{subfigure}
\caption{Qualitative comparison of FreeTimeGS and TSOG, for Neural 3D sequence Cut Roasted Beef, frame \#77.} 
\label{fig:qualitative_comparison_tsog}
\vspace{-0.5cm}
\end{figure}

\begin{table}[h]
\caption{Comparison between PLYs sequences and TSOG. Size is in MBs.}
\label{tab:compression_results_sequence}
\centering
\resizebox{0.49\textwidth}{!}{%
\renewcommand{\arraystretch}{1.2}
\begin{tabular}{l l c c c c c c c c c c}
\toprule
\textbf{Dataset} & \textbf{Scene}
& \multicolumn{3}{c}{\textbf{PLYs sequence}}
& \multicolumn{3}{c}{\textbf{TSOG}} \\
\cmidrule(lr){3-5} \cmidrule(lr){6-8}
&
& \textbf{PSNR} & \textbf{SSIM} & \textbf{Size} 
& \textbf{PSNR} & \textbf{SSIM} & \textbf{Size} \\
\midrule
VV & 001\_0\_seq0
& 23.9443 & 0.8678 & 5640
& 23.5733 & 0.8671 & 152 \\

VV & 012\_0\_seq0
& 25.8669 & 0.9102 & 5350
& 25.8733 & 0.9101 & 165 \\

Neural 3D & Coffee Martini
& 23.0219 & 0.84612 & 4690
& 23.8735 & 0.8622 & 142 \\

Neural 3D & Cook Spinach
& 25.2826 & 0.8926  & 4500
& 25.7933 & 0.9003  & 141 \\

Neural 3D & Cut Roasted Beef
& 24.8606 & 0.8983 & 4590
& 25.2457 & 0.9062 & 144 \\

Neural 3D & Flame Steak
& 25.8930 & 0.9174 & 4380
& 26.2454 & 0.9230 & 147 \\

Neural 3D & Sear Steak
& 26.8553 & 0.9133 & 4550
& 27.6302 & 0.9241 & 155 \\
\bottomrule
\end{tabular}}
\vspace{-0.25cm}
\end{table}

Table~\ref{tab:compression_results_sequence} reports the quantitative comparison between PLYs sequence and TSOG file formats. 
Based on our results, TSOG reduces storage requirements by 96.88\%, and increases PSNR and SSIM by 1.43\% and 0.76\% respectively, across all scenes. 
The PSNR differences range from -0.37 to 0.85 dB and SSIM from -0.0007 to 0.0161.
We attribute the quality gains to the fact that TSOG tends to regularize the representation, which can supress small inconsistencies or noise present in the PLYs sequence.
The average decoding time for the PLYs sequence is 16.15 and 20.72~sec in VV and Neural 3D datasets, whereas for the TSOG the decoding times are 8.04 and 7.15~sec, representing a reduction of 50.25\% and 65.5\%, respectively. 
The average encoding time for TSOG is 1066.5 and 1122.4~sec in VV and Neural 3D dataset, respectively.
In Figure~\ref{fig:qualitative_comparison_tsog}, exemplary frames stored in PLYs sequence and TSOG are shown.

\begin{figure}[t]
\centering
\begin{subfigure}{0.5\textwidth}
\centering
\begin{subfigure}{0.49\textwidth}
  \centering
  \includegraphics[width=\linewidth]{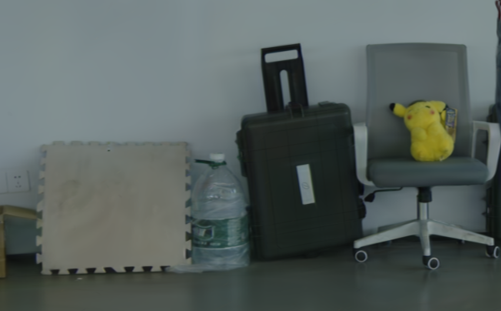}
  \caption{16-bit temporal means}
  \label{fig:q_ply_001_crop}
\end{subfigure}
\begin{subfigure}{0.49\textwidth}
  \centering
  \includegraphics[width=\linewidth]{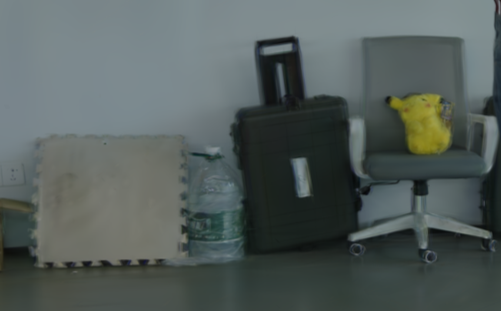}
  \caption{8-bit temporal means}
  \label{fig:q_tsog_001_crop}
\end{subfigure}
\end{subfigure}
\caption{Qualitative comparison of TSOG representations using 16-bit and 8-bit temporal means, for VV sequence 001\_1\_seq0, frame \#208, cropped region.} 
\label{fig:qualitative_comparison_tsog_temporal_means}
\vspace{-0.5cm}
\end{figure}

\subsection{FreeTimeGS Representation}
\label{ssec:ftgs}
In this experiment, FreeTimeGS represents the baseline, with 4D Gaussians parameterized by continuous temporal attributes. 
To convert FreeTimeGS content in the TSOG file format, the base values of position attributes are stored as static attributes. 
Higher-order coefficients describing the linear evolution of position values are represented by per-Gaussian velocity vectors, and stored as a dynamic attribute in the temporal means image, specified under the \texttt{temporal} structure. 
The timeline attribute is configured with \texttt{type} set to 2, mapping the \textit{center} and \textit{scale} to the mean and standard deviation values used in FreeTimeGS's exponential parameterization of a Gaussian's temporal opacity. 

For the VV~\cite{vvc2025siggraph} dataset, we use the full 300 frames of each scene, whereas for the Neural 3D dataset, we use the first 150 trained frames.
As part of this experiment, we evaluate 16-bit and 8-bit precisions for the temporal means.

\begin{table}[h]
\caption{Comparison between PLY extended for FreeTimeGS and TSOG using 16-bit temporal means. Size is in MBs.}
\label{tab:compression_results}
\centering
\resizebox{0.49\textwidth}{!}{%
\renewcommand{\arraystretch}{1.2}
\begin{tabular}{l l c c c c c c c c c c c c c c c}
\toprule
\textbf{Dataset} & \textbf{Scene}
& \multicolumn{3}{c}{\textbf{PLY extended}}
& \multicolumn{3}{c}{\textbf{TSOG}} \\
\cmidrule(lr){3-5} \cmidrule(lr){6-8} 
&
& \textbf{PSNR} & \textbf{SSIM} & \textbf{Size}
& \textbf{PSNR} & \textbf{SSIM} & \textbf{Size} \\
\midrule
VV & 001\_0\_seq0
& 23.9443 & 0.8678 & 1130
& 23.5194  & 0.8673 & 95.7 \\

VV & 012\_0\_seq0
& 26.5832  & 0.9197 & 1120
& 26.4534 & 0.9167 & 96.3  \\

Neural 3D & Coffee Martini
& 23.6124 & 0.8626 & 1085
& 23.9534 & 0.8670 & 85.3  \\

Neural 3D & Cook Spinach
& 25.6740 & 0.8998 & 1082
& 26.0419 & 0.9032 & 86.9  \\

Neural 3D & Cut Roasted Beef
& 25.0195 & 0.9051 & 1084
& 25.3523 & 0.9075 & 85.9  \\

Neural 3D & Flame Steak
& 25.7726 & 0.9209 & 1079
& 26.0240 & 0.9224 & 87.1  \\

Neural 3D & Sear Steak
& 27.5206 & 0.9229 & 1080
& 27.9311 & 0.9260 & 87.3 \\
\bottomrule
\end{tabular}}
\vspace{-0.25cm}
\end{table}

Table~\ref{tab:compression_results} reports the quantitative comparison between FreeTimeGS, stored in a model-specific PLY extended format, and the TSOG file format. 
TSOG reduces the file size on average by 91.85\%, with PSNR and SSIM gains of 0.63\% and 0.18\%, across all scenes, when using 16-bit temporal means. 
The PSNR differences range from -0.42 to 0.41 dB and SSIM from -0.003 to 0.004.
The average decoding time for FreeTimeGS is 4.43 and 4.04 sec in VV and Neural 3D datasets, whereas for TSOG the decoding times are 1.76 and 1.92 sec, representing a decrease of 60.27\% and 52.38\%, respectively. 
The average encoding time for TSOG is 266.24 and 259.03 sec in VV and Neural 3D datasets, respectively.
For 8-bit temporal means, the file size reductions increase to 93.03\%, at substantial quality degradations in PSNR and SSIM of 5.09\% and 5.98\%, and marginal saves in average encoding and decoding times of 2.76\% and 3.88\%, respectively, across all scenes, favoring the usage of higher bit depth.
In Figure~\ref{fig:qualitative_comparison_tsog}, exemplary frames using FreeTimeGS and TSOG with 16-bit temporal means are shown, while in Figure~\ref{fig:qualitative_comparison_tsog_temporal_means}, cropped regions are magnified to showcase the blurring artifacts introduced due to quantization errors using 8-bit temporal means.

\section{Demonstration}
\label{sec:demo}
\begin{figure}[t]
\centering
  \centering
  \includegraphics[width=0.78\linewidth]{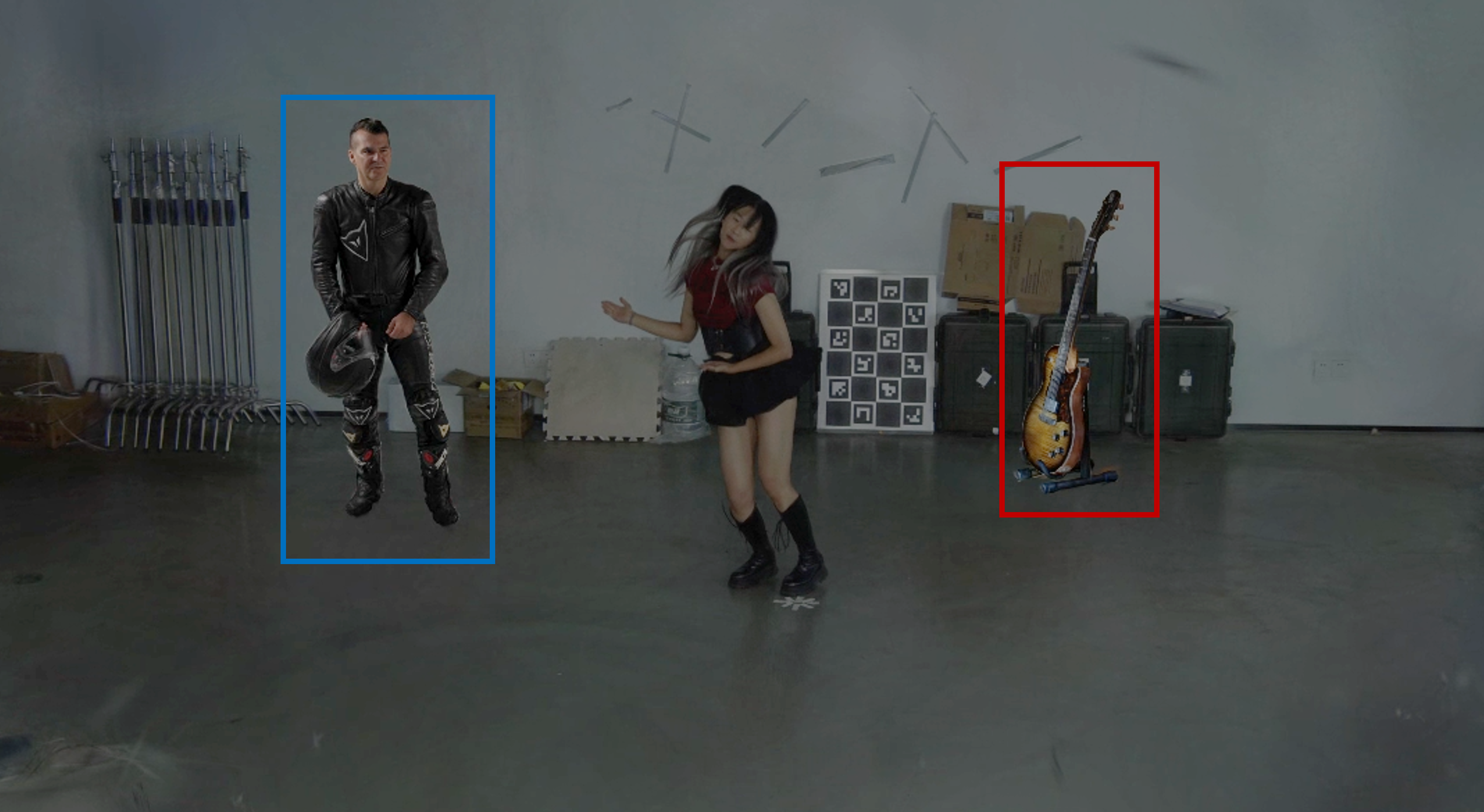}
\caption{Selected frame from a 4DGS content, superimposed with two external objects in the same scene, highlighted with blue and red bounding boxes.} 
\label{fig:playcanvas}
\vspace{-0.4cm}
\end{figure}

We demonstrate the feasibility of TSOG representation format by integrating it into PlayCanvas~\cite{playcanvas_2026}, which is a widely used browser-based engine for real-time 3DGS visualization.  
We extend PlayCanvas to parse and decode TSOG files, enabling dynamic playback of 4DGS content through index-aligned attribute images and temporal parameterizations. 
The implementation is designed to be lightweight and reusable. 
Figure~\ref{fig:playcanvas} shows a selected frame from the playback of a 4DGS scene, highlighting external objects added to the scene with two bounding boxes, to illustrate the capabilities of PlayCanvas and the types of experiences enabled.

\section{Conclusions}
\label{sec:conclusions}
In this paper, we introduced TSOG, an extension of SOG~\cite{sog_2026}, that defines a format for representing, storing, and delivering 4DGS content in a structured and interoperable manner. 
TSOG adds a per-Gaussian timeline, defined on a shared scene time axis, and supports temporal parameterization of attributes while preserving the key SOG property that each Gaussian maps to the same pixel location across all attribute images. 
This design ensures the TSOG format remains model-agnostic and backward compatible with SOG decoding pipelines. 
In our experiments, the TSOG file format reduces file size by over 90\% compared to PLY-based baselines, at a similar visual quality.
Future work may extend TSOG to cover a broader set of temporal parameterization models and evaluate its performance under deployment-oriented conditions, such as streaming and web-based playback.

\newpage
\bibliographystyle{IEEEbib}
\bibliography{strings,refs}

\end{document}